\documentclass[twocolumn]{aastex631}
\usepackage{subfigure}
\usepackage{amsmath}
\usepackage{esint}
\usepackage{color, colortbl}
\definecolor{Gray}{gray}{0.9}
\shorttitle{Schwarzschild Modeling of Barred Galaxies}
\shortauthors{B. Tahmasebzadeh et al.}
\graphicspath{{./}{figures/}}

\begin{document}

\title{Orbit-Superposition Dynamical Modeling of  Barred Galaxies}

\author[0000-0002-1584-2281]{Behzad Tahmasebzadeh} \thanks{behzadtahmaseb@gmail.com;}
\affiliation{Shanghai Astronomical Observatory, Chinese Academy of Sciences, 80 Nandan Road, Shanghai 200030, China}
\affiliation{Department of Astronomy and Space Sciences, University of Chinese Academy
of Sciences, 19A Yuquan Road, Beijing 100049, China}

\author[0000-0002-8005-0870]{Ling Zhu}\thanks{Corr authors: lzhu@shao.ac.cn;} 
\affiliation{Shanghai Astronomical Observatory, Chinese Academy of Sciences, 80 Nandan Road, Shanghai 200030, China}

\author[0000-0001-5604-1643]{Juntai Shen}\thanks{jtshen@sjtu.edu.cn.} 
\affiliation{Department of Astronomy, School of Physics and Astronomy, Shanghai Jiao Tong University, 800 Dongchuan Road, Shanghai 200240, China}
\affiliation{Key Laboratory for Particle Astrophysics and Cosmology (MOE) / Shanghai Key Laboratory for Particle Physics and Cosmology, Shanghai 200240, China}

\author[0000-0003-3333-0033]{Ortwin Gerhard}
\affiliation{Max-Planck-Institut f{\"u}r Extraterrestrische Physik, Gie{\ss}enbachstra{\ss}e 1, 85748 Garching, Germany}

\author[0000-0003-4546-7731]{Glenn van de Ven}
\affiliation{Department of Astrophysics, University of Vienna, T\"urkenschanzstra{\ss}e 17, 1180 Vienna, Austria}



\begin{abstract}
Barred structures are important in understanding galaxy evolution, but they were not included explicitly in most dynamical models for nearby galaxies due to their complicated morphological and kinematic properties. 
We modify the triaxial orbit-superposition Schwarzschild implementation by  van den Bosch et al. to include barred structures explicitly. The gravitational potential is a combination of a spherical dark matter halo and stellar mass; with the 3D stellar density distribution deprojected from the observed 2D image using a two-component de-projection method, including an axisymmetric disk and a triaxial barred bulge. We consider figure rotation of the galaxy with the bar pattern speed as a free parameter. We validate the method by applying it to a mock galaxy with integral field unit (IFU) data created from an \textit{N}-body simulation with a boxy/peanut or X-shaped bar. Our model fits the observed 2D surface density and all kinematic features well. The bar pattern speed is recovered well with a relative uncertainty smaller than $10 \%$. Based on the internal stellar orbit distribution of the model, we decompose the galaxy into an X-shaped bar, a boxy bulge, a vertically extended structure and a disk, and demonstrate that our model recovers these structures generally well, similar to the true structures in the \textit{N}-body simulation. Our method provides a realistic way of modeling the bar structure explicitly for nearby barred galaxies with IFU observations.
\end{abstract}

\keywords{Barred spiral galaxies (136); Galaxy dynamics (591); Galaxy structure (622)}


\section{Introduction}
About half of disk galaxies have barred structures with a variety morphological and  kinematic appearances \cite[]{Eskridge.2000, Erwin.2018}. When the galaxy is face-on or moderately inclined, bars can appear as nonaxisymmetric perturbations in the surface density. In the case of edge-on or highly inclined galaxies, bars can be revealed by kinematic imprint, e.g., a positive correlation between mean velocity and the third Gauss-Hermite moment $ h_{3} $ \cite[]{2005.Bureau, 2018.Li}. Bars can grow in the vertical direction and display a boxy/peanut or X-shaped (hereafter BP/X) structure in edge-on views. BP/X bulges are found in numerous \textit{N}-body simulations and observations \cite[]{Combes.1981,Raha.1991, Ltticke.2000, Erwin.2017}. Bars play an important role in  galaxy secular evolution since they can redistribute the angular momentum and energy of the disk materials \cite[]{1998.Debattista, 2003.Athanassoula, KK.2004, 2011.Gadotti}.
\par
Observations with integral field unit (IFU) instruments have significantly improved our understanding of galaxies. In the past decades, IFU surveys such as CALIFA \cite[]{CALIFA.2012}, SAMI \cite[]{SAMI.2012}, and MaNGA \cite[]{MANGA.2015}, provided kinematic maps of thousands of nearby galaxies. Recently, a sample of 24 nearby barred galaxies have been observed with MUSE on the Very Large Telescope as part of the TIMER project \cite[]{TIMER.2020}, which shows in great detail the stellar kinematics and populations of the barred structures, and reveals the existence of inner structures like nuclear rings and inner disks. The IFU data are
powerful in probing the formation of different structures, but the information is still integrated along the line of sight. To further uncover the galaxies' underlying luminous and dark mass
distributions, as well as the galaxies' 3D shape and internal orbital structures, we need dynamical models. The pattern speed of barred galaxies is another important parameter could be accurately obtained by dynamical modeling.
\par
The \cite{Schwarzschild.1979} orbit-superposition method is a
practical approach for constructing dynamical models for stellar
systems without ad hoc assumptions on the underlying distribution
functions (DFs). 
The method was used for theoretic studies in some early works, such as studying the importance of different orbit families in stellar systems \cite[]{1993.Schwarzschild, 1996.Merritt, Vasiliev.2013}, and generation of initial equilibrium conditions for \textit{N}-body simulations \cite[]{Vasiliev.2015}. The method is potentially flexible for complicated galaxy structures. 
\par
Constraining the Schwarzschild models to fit photometric and kinematic observational data was started by \cite{Pfenniger.1984} and \cite{Richstone.1984, Richstone.1985}.  Currently, there are several commonly used  implementations of the Schwarzschild's orbit-superposition method in different geometries, including those for spherical systems \cite[]{Richstone.1985, 2013.Breddels, 2017.Kowalczyk}, axisymmetric systems \cite[]{1999.Cretton, Gebhardt.2000, 2004.Valluri, Cappellari.2006, Thomas.2007, Saglia.2016, Thater.2019, Thater.2022}, and triaxial systems \cite[]{bosch.2008, Jin.2019, Neureiter.2021}. In particular, the \cite{bosch.2008} triaxial orbit-superposition code (hereafter \texttt{VdB08}) has been widely used in measuring the central supermassive BH mass \cite[e.g.,][]{Bosch.2010, Walsh.2012, Seth.2014, Walsh.2015, Ahn.2018, Feldmeier.2017, Quenneville.2022}, probing the underlying luminous and dark mass distribution and stellar orbit distribution for large sample of galaxies across the Hubble types from CALIFA \citep{ling.1018}, MaNGA \citep{Jin.2020} and SAMI \citep{Giulia.2022}, and studying formation history of galaxies by tagging orbits with stellar population properties \cite[]{Poci2019,Zhu.2020,Poci.2021,Zhu.2022}. A new implementation of \texttt{VdB08} code named DYNAMITE  has been publicly released with some new features \cite[]{Jethwa.2020,Sabine.2022}. However, rotating barred structures are not included explicitly in most of the existing Schwarzschild models for external galaxies. 
\par
The barred structure of the Milky Way has been carefully modeled in several works. In some early works, 3D stellar density distributions deprojected from COBE photometry were used as model input of Schwarzschild dynamical models  \citep{Zhao.1996, Hafner.2000}. In the recent works, intrinsic 3D stellar density from \textit{N}-body simulations were used as model input, for either an orbit-based Schwarzschild model \citep{Wang.2013} or particle-based made-to-measure (M2M) method \citep{Long2013}.
The input 3D density distribution from \textit{N}-body simulations could be adjusted according to observed stellar number densities of the Milky Way, thus guaranteed to be a good approach of the real galaxy \cite[]{Portail.2017a}.
These works obtained pattern speed of $\sim 40$ km s$^{-1}$ kpc$^{-1}$ for the Milky Way bar, consistent with each other. Particles in the Made-to-Measure method have been further tagged with metallicity to study the stellar population of the bar structure \citep{Portail.2017}.
\par
Dynamical modeling of external barred galaxies is still at the early stage. 
\cite{Pfenniger.1984} built a 2D dynamical Schwarzschild model for a barred galaxy and compared the best-fitting model to the velocity field of NGC 936.  \cite{blana.2018} made a triaxial bulge/bar/disk M2M model for M31, taking an \textit{N}-body simulation that generally matches the bulge properties of M31 as an initial condition of the M2M algorithm. A bar has been included in the recently developed Schwarzschild  FORSTAND code \cite[]{Vasiliev.2019c}, which is applied to mock data created from a simulation by using its real 3D density distribution. 
\par
Estimating the 3D density distribution is a crucial step before we can create proper dynamical models for a real external barred galaxy. In previous work \cite[]{Behzad.2021}, we have developed a two-component de-projection method to infer the 3D density distribution of an external barred galaxy from its observed 2D image, by including an axisymmetric disk and a triaxial (mostly prolate) barred bulge.
In this paper, we employ this method to obtain the 3D stellar density distribution, which will be used in constructing the stellar mass distribution in the gravitational potential and in constraining the internal 3D density distribution of the dynamical model. We further modify the \texttt{VdB08} triaxial Schwarzschild code by including an appropriate orbits sampling, and considering the bar figure rotation. We validate the new method by applying it to mock IFU data created from an \textit{N}-body simulation with a BP/X bulge.
\par
The paper is organized as follows. We describe the modeling construction and technical details of the code modification in Section \ref{S:method}. We validate the code by testing against the mock data and discuss the results compared to the original simulation in Section \ref{S:mock}. We summarize and conclude in Section \ref{S:con}.

\section{Model Construction}\label{S:method}
This section gives a detailed description of the model construction for a barred galaxy with modifications to the \texttt{VdB08} model. There are three major steps to create a model: (1) constructing the gravitational potential, (2) orbit sampling and integration, and (3) finding the weights of the orbits by fitting to the observational data.  We updated the de-projection procedure to allow a triaxial (mostly prolate) bar/bulge embedded within an axisymmetric disk in the stellar contribution of the gravitational potential and in the 3D luminosity distribution of the tracer. We include a non-zero figure rotation of the galaxy. The model is constructed in the bar co-rotating frame, then the orbit sampling, calculation and storage are updated accordingly. The procedure of orbit weighing are kept generally the same as in \texttt{VdB08}. 

\subsection{Gravitational Potential}\label{S:pot}
The gravitational potential is a combination of stellar mass and dark
matter. We do not include a central black hole (BH) here as the BH sphere of influence is not resolved by
the kinematic data we are using, thus the BH has no effect on our model fitting. However, a BH could be included straightforwardly if high resolution data are available.

\subsubsection{Stellar Mass}
Photometric images trace the stellar light of the galaxy, but on the 2D sky plane. In \cite{Behzad.2021}, we developed a two-component de-projection method for barred galaxies which is used here to obtain the intrinsic 3D luminosity distribution. In this approach, the image of a barred galaxy is first decomposed into a disk and an elliptical bulge by GALFIT  \cite[]{Peng.2010}, we subtract the disk from the original image, which leads to a residual barred bulge. We then apply the multi-Gaussian expansion (MGE) fit to the disk and the residual barred bulge separately, and deproject them individually by assuming the disk is axisymmetric while the barred bulge is triaxial. Note that we use the residual barred bulge, which allows us to capture the triaxiality better than the fitted elliptical bulge.
By combining the barred bulge and the disk, we obtain the 3D density distribution of the barred galaxy. In the following, we briefly review some technical aspects of our method.
\par
The MGE fitting to the surface brightness (in units of $\mathrm{L_{sun}} \mathrm{pc}^{-2}$) of the disk and the barred
bulge is performed separately following \citep[]{cappellari.2002}
\begin{equation}
\label{MGESB}
\mathbf{\Sigma}\left(R^{\prime}, \theta^{\prime}\right)=\sum\limits_{\substack{j=1 }}^N \frac{L_{j}}{2\pi \sigma'^{2}_{j}q'_{j}}\exp \left[-\frac{1}{2\sigma'^{2}_{j}} \left( x_{j}^{\prime 2}+\frac{y_{j}^{\prime 2}}{q'^{2}_{j}} \right) \right],
\end{equation}
where $L_{j}$ is the total luminosity, $q_{j}^{\prime}$ is the projected flattening, and $\sigma_{j}^{\prime}$ is the scale length along the projected major axis
of each Gaussian component $j=1 \dots N$. Furthermore, $x_{j}^{\prime}$ and $y_{j}^{\prime}$ are related to the polar coordinates in the sky plane $ (R^{\prime}, \theta^{\prime}) $ via
\begin{equation}
{x_{j}^{\prime}=R^{\prime} \sin \left(\theta^{\prime}-\psi_{j}^{\prime} \right)}, \hspace{0.5cm} {y_{j}^{\prime}=R^{\prime} \cos \left(\theta^{\prime}-\psi_{j}^{\prime} \right)},
\end{equation}
The position angle of $\psi_{j}^{\prime}$ is measured counterclockwise from the major axis of each Gaussian component to the $y^{\prime}$-axis. We denote:
\begin{equation}
\label{pstw}
\psi_{j}^{\prime}=\psi+\Delta \psi_{j}^{\prime}
\end{equation}
where $ \Delta \psi_{j}^{\prime}$ is the difference of $\psi_{j}^{\prime}$ to the global position angle $\psi$ of the object. $ \Delta \psi_{j}^{\prime}$ is thus the isophotal twist of each Gaussian, and it can be measured directly during the MGE fitting.
\par 
\par
Then, we deproject the 2D MGE surface brightness to obtain the 3D MGE luminosity density 
\begin{equation}\label{MGESB2}
\begin{split}
\rho(x,y,z)= \sum\limits_{\substack{j=0 }}^N & \frac{L_{j}}{(\sigma_{j} \sqrt{2\pi })^{3} q_{j} p_{j}}  \\
& \times \exp \left[-\frac{1}{2\sigma^{2}_{j}} \left( x^{2}+\frac{y^{2}}{p^{2}_{j}}+\frac{z^{2}}{q_{j}^{2}} \right) \right], 
\end{split}
\end{equation}
where $p_{j}$ and $q_{j}$ are the intermediate-to-long and short-to-long axis rations of the Gaussian component $j$.
\par 
For the disk and the barred bulge separately, by fitting
Eq. (\ref{MGESB}) to the surface brightness, we have had the
parameters $(L_{j}, q'_{j} ,\sigma '_{j}, \Delta \psi_{j}^{\prime} ) $
of the 2D Gaussians. We use three viewing angles $
  (\theta,\varphi,\psi) $ to define the orientation of a projected
  system, $ \theta $ and $ \varphi $ indicate the orientation of the
  line-of-sight with respect to the principal axes of the object. For
  example, projections along the intrinsic major, intermediate, and
  minor axes correspond to $(\theta=90^{\circ},\varphi=0^{\circ}) $,
  $(\theta=90^{\circ},\varphi=90^{\circ}) $ and
 ($\theta=0^{\circ}$, $\varphi$ irrelevant),
  respectively. The rotation of the object around the line of sight in
  the projected plane is specified by $ \psi $  (see Fig. 2 in \cite{Zeeuw.1989}). 
Given a set of viewing angles $ (\theta, \varphi, \psi) $, the intrinsic quantities $ (\sigma_{j}, p_{j}, q_{j}) $ can be obtained analytically using the parameters measured from the 2D Gaussians (see Eqs. (7-9) in  \texttt{VdB08}). All Gaussians should have the same viewing angles $(\theta, \varphi, \psi) $ for a rigid body, so the allowed viewing angles for one component are the intersection of allowed orientations of all the Gaussians fitting that component. 

We thus have three viewing angles $(\theta_{\rm disk}, \varphi_{\rm disk}, \psi_{\rm disk}) $ for the disk and $(\theta_{\rm bar}, \varphi_{\rm bar}, \psi_{\rm bar}) $ for the barred bulge. 
We consider the disk is an axisymmetric oblate system with the major axis aligned with the $ x^{\prime} $ axis of the model so that $\psi_{\mathrm{disk}} = 90^{\circ}$, and $ \Delta \psi_{j}^{\prime} = 0 $ for all Gaussians, while $\varphi _{\rm disk}$ is irrelevant. We are left with the inclination angle of the disk $\theta_{\rm disk}$ as a free parameter, with its lower limit constrained by $\cos( \theta_{\rm disk} )^2< q'^2_{\rm min} $ where $q'_{\rm min}$ is the flattest Gaussian in the MGE fitting to disk. 

For the barred bulge, the different Gaussians are allowed to have
different isophotal twists $\Delta \psi_{j}^{\prime}$. We measure the
twists of Gaussian components in the barred bulge with respect to the
disk major axis in the observational plane, we thus have $ \Delta \psi_{j}^{\prime}  =
\psi_{j}^{\prime}-\psi_{\mathrm{disk}} $, and $\psi_{\rm bar} = \psi_{\rm disk}$, with the real information of bar position angle included in $\Delta \psi_{j}^{\prime}$.
When combining the bar and disk together, we enforce the major axis of the bar to be aligned within the disk plane, and the inclination angle constrained to be $\theta_{\rm bar} = \theta_{\rm disk}$, while the angle $\varphi_{\rm bar}$ is left free.
\par
We thus have two viewing angles as free parameters in the de-projection: $\theta_{\rm disk}$ and $\varphi_{\rm bar}$,
which will be just denoted as $\theta$ and
$\varphi$ in what follows. Once we have the 3D luminosity density distribution, we further multiply it with a stellar mass-to-light ratio $M_*/L$ to arrive at the 3D stellar mass distribution. The stellar mass-to-light ratio $M_*/L$, which we assume here to be constant, is the third free parameter in the mass model.
\subsubsection{Dark Matter}
We consider a spherical Navarro–Frenk–White (NFW) halo \cite[]{NFW.1996} to represent a DM distribution with 
an enclosed mass profile expressed as 
\begin{equation}\label{DM:NFW}
M(<r)=M_{200} g(c)\left[\ln \left(1+c r / r_{200}\right)-\frac{c r / r_{200}}{1+c r / r_{200}}\right]
\end{equation}
where $g(c)=[\ln (1+c)-c /(1+c)]^{-1}$ and $ c $ is the concentration of the DM halo.
 $M_{200}=\frac{4}{3} \pi 200 \rho_{\mathrm{c}} r_{200}^{3}$ indicates the virial mass, which is defined as the mass within the virial radius $r_{200}$. The critical density
is adopted as $\rho_{\mathrm{c}}=1.37 \times 10^{-7} M_{\odot} \mathrm{pc}^{-3}$. Thus, two free parameters remain in the NFW halo: the concentration $c$ and the virial mass $ M_{200} $. 
\par
Since the data are not extended to sufficient large radius, we cannot constrain $c$ and $M_{200}$ at the same time. Therefore, we fix $c$ based on the relation %
\begin{equation}\label{DM:c}
\log _{10} c=0.905-0.101 \log _{10}\left(M_{200} /\left[10^{12} h^{-1} M_{\odot}\right]\right)
\end{equation}
inferred from galaxy simulations \citep[]{Dutton.2014} with $h = 0.671$ \citep[]{plank.2014}. 
We thus only have one free parameter, $ M_{200} $, in the DM mass distribution. The contribution of DM within the radius of the outermost kinematic aperture could be well constrained by the model, which is directly related to $M_{200}$ taken the assumption of equation (6). 
\subsubsection{Figure Rotation}
For dynamical modeling of a barred galaxy, we consider a figure rotation of the galaxy, and the gravitational potential is stationary in the rotating frame ($x$ is the bar major axis direction). We integrate the orbits in the rotating frame in which Jacobi energy ($E_{J}$) is an integral of motion:
\begin{equation}
E_{\mathrm{J}}=\frac{1}{2}|\dot{\vec{r}}|^{2}+\Phi-\frac{1}{2}\left|\vec{\mathbf{\Omega}} \times \vec{r}\right|^{2}
\end{equation}
where $\vec{r}$ and $\dot{\vec{r}}$ are three-dimensional spatial and velocity vectors in the rotating frame, respectively. $\vec{\mathbf{\Omega}}$ indicates the angular velocity vector. We check  $E_{J}$ to be conserved during the orbit integration. 

The equations of motion in the rotating frame are
\begin{equation}
\ddot{\vec{r}} =-\nabla \Phi-2\left(\vec{\mathbf{\Omega}} \times \dot{\vec{r}}\right)-\vec{\mathbf{\Omega}} \times\left(\vec{\mathbf{\Omega}} \times \vec{r}\right) 
\end{equation}
 where the second and the third terms are the components of Coriolis force and the centrifugal force, respectively.

Using the Cartesian coordinates and adopting the counterclockwise motion about the $z$-axis with $\Omega>0$, the equations of motion can be expressed as
\begin{equation}
\label{eqn:em2}
\begin{split}{}
\dot{x}=v_x+\Omega y, \hspace{0.5cm} & \dot{v}_x=-\frac{\partial \Phi}{\partial x}+\Omega v_y \\
\dot{y}=v_y-\Omega x, \hspace{0.5cm} & \dot{v}_y=-\frac{\partial \Phi}{\partial y}-\Omega v_x \\
\dot{z}=v_z, \hspace{0.5cm} & \dot{v}_z=-\frac{\partial \Phi}{\partial z}
\end{split}
\end{equation}
where $v_x$, $v_y$, and $v_z$ are the velocities in the inertial frame, but instantaneously in the coordinate that $x$ aligns with the bar.
We record the information of $x,y,$ and $z$ and $v_x$, $v_y$, and $v_z$ during the orbit integration to produce kinematic data cubes comparing with observations.

The bar pattern speed $\Omega$ is left as a free parameter.
Finally, we have five free so-called hyperparameters in the model: inclination $\theta$, bar azimuthal angle $\varphi$, stellar mass-to-light ratio $M_*/L$, DM virial mass $M_{\rm 200}$, and pattern speed $\Omega$.

\subsection{Orbit Sampling and Integration}
\subsubsection{Initial Conditions}
We sample the initial conditions of orbits in the $ x-z $ plane, using
the properties of separable models following \texttt{VdB08}. In a
separable model, the tube orbits (except the shell orbits in which the outer and inner radial turning points coincide) will pass through the $ x-z $ plane perpendicularly twice above $z>0$. So sampling of the whole $ x-z $ plane is not necessary. We sample the orbital energy $ E $ through a logarithmic grid in radius; each energy is linked to a grid radius $ r_{i} $ by calculating the potential at the position $ (x, y, z) = (r_{i}, 0, 0)$. Then for each energy, the starting point $ (x, z) $ is selected from a linear open polar grid of the $ (R, \phi) $ in between the location of the shell orbits and the equipotential surface with zero velocity of this energy, where $R=\sqrt{x^2 + z^2}$ and $\phi=\arctan(x/z)$ (the gray area in Fig 2 of \texttt{VdB08}). Note that the location of the shell orbits curves are found iteratively to avoid double counting of initial starting points. This is done by launching orbits at different radii by fixing $\phi$, until the orbit width become minimal. An alternative scheme is to place positions at the intermediate $y-$axis and velocities in the $v_x-v_z$ plane, where all four orbit families can be sampled \cite[]{Schwarzschild.1982,Deibel.2011}. The number of initial conditions we sampled is $n_{E} \times n_{R} \times n_{\phi} $. The starting point on the $ x-z $ plane is set as $ y = 0$, $v_{x} = v_{z} = 0 $, and $ v_{y}=\sqrt{2[E-\Phi(x, 0, z)]}$.
The initial starting points are sampled in inertial frame, and converted to velocities in the rotating frame for the orbit integration following equation~\ref{eqn:em2}.
\par
We use an orbit-dithering approach to impose the smoothness of orbit-superposition models. It increases the number of starting points $n_{\mathrm{dith}}$ times at each direction of $(E, R, \phi)$. It leads to $n_{\mathrm{dith}}^{3}$ orbits per orbital bundle. The properties of orbits in each bundle are coadded, and all observable data cubes are averaged over each bundle.
\par
To cover the different types of orbits supporting the bar, we sample a large number of starting points across the three integrals with $(n_{E} \times
n_{R} \times n_{\phi}) = (40 \times 20 \times 10)$, and we adopt the dithering number to be 3, so each orbital bundle contains $27$ orbits with close starting points.
\par
In \texttt{VdB08} for a stationary potential, the retrograde orbits are not integrated individually, but taking into account only in the fitting procedure by flipping the sign of $v_{y}$ from the prograde orbits. In a triaxial potential with nonzero pattern speed, the retrograde orbits will behave differently from the prograde orbits, we thus have to integrate them individually. We sample another set of initial conditions at the same energy intervals and $x-z$ plane, but with $ v_{y}=-\sqrt{2[E-\Phi(x, 0, z)]} $. The number of starting points are chosen exactly the same as the prograde orbits. 
\par
In the above orbit libraries sampled from the $x-z$ plane, box orbits are abundant in the inner regions, but become rare at large radius. Another set of box orbit library was sampled through equipotential curves in stationary start space in \texttt{VdB08}, to increase the number of box orbits, especially at large radii, which is important for modeling the triaxial giant elliptical galaxies.
But including this box orbit library or not does not make any noticeable difference in modeling spiral galaxies, which are dominated by disks in the outer regions. We thus do not include it in our model. In practice, we will show in Section \ref{S:ron_alz} that we already have all the typical orbit families of a barred spiral galaxy with the orbit libraries sampled in the $x-z$ plane.

\subsubsection{Orbit Integration and Storage}
We integrate each orbit for 200 $t_{\rm dyn}$ ($t_{\rm dyn}$ is defined as the period of a closed elliptical orbit with the same energy), and store 50,000 points per orbit with equal time intervals. We start the integration with target relative accuracy of $10^{-5}$ in energy conservation. After the integration of each orbit, its conservation of Jacobi energy $E_J$ is checked. If the $E_J$ at the end has changed more than 1\% of the initial value, the orbit is re-integrated with a higher target energy accuracy. Following \texttt{VdB08}, we use the DOP853 explicit Runga-Kutta integrator.
\par
The stationary, nonrotating galaxies are symmetric in the three principal planes. In \texttt{VdB08}, all orbit properties were calculated in only one octant; the properties in the other octants were symmetrized by an eight-fold symmetry, as described in \texttt{VdB08} and with a bug recently reported by \cite{Quenneville.2022} and corrected in the latest version
of publicly released DYNAMITE code \cite[]{Sabine.2022}.
\par
In modeling a barred galaxy with figure rotation, the symmetry in the $x-y$ plane is broken and thus restricted to a four-fold symmetrization. For each orbit that we only sample the initial starting point in one octant, we obtain the other three mirror orbits by flipping the signs of positions and
velocities following Table 1, which is revised from \cite{Sabine.2022}.  We combine four mirror orbits together and treat it as one single orbit in the model fitting.
\par
In the bar rotating frame, long-axis tube orbits tilt due to Coriolis force that could break the triaxial symmetry \cite[]{Valluri.2016}. To still keep the symmetry, here we enforce long-axis tube orbits to flip the signs in the same way as box orbits. This eliminates the possibility of net rotation about the major axis, which is responsible for kinematic twists in a nonrotating system, while this may not be a problem for a barred galaxy. We studied the orbital structures of different \textit{N}-body bars, and do not detect appreciable streaming motion about the $x$-axis in \textit{N}-body bars. The percentage of long-axis tube orbits in all models are less than $2 \%$ and with similar contributions of positive and negative $L_{x}$ (Tahmasebzadeh et al. 2022 in preparation). Our choice for the symmetry of long-axis tube orbits should be feasible for modeling of barred galaxies. In these four mirrors, short-axis tube orbits also flip the signs in the same way as box orbits \cite[]{Sabine.2022}. Hence, the symmetrization pattern is identical for all orbit families in a barred galaxy model with figure rotation.  

\par
Although we perform orbit integration in the bar corotating frame, the kinematics in the inertial frame $v_x,v_y,$ and $v_z$ are stored for the orbits, in the coordinate that $x$-axis instantaneously aligns with the bar at the corresponding moment of time.
Key information of each orbit combined with its three mirrors are stored in two ways: 
(1) We store information projected to the 2D observational plane and in the same observational apertures as the real data, including the surface brightness and full line of sight velocity distribution (LOSVD) stored in a  histogram. 
(2) We store 3D intrinsic properties in 3D spherical grids, including the intrinsic 3D density distribution for fitting the 3D density distribution of the galaxy and 3D kinematic information for later analyses. In our spherical 3D grids ($r_{\rm grid}, \theta_{\rm grid}, \phi_{\rm grid}$), the radial grids are sampled logarithmic from the inner of $10^{-2}$ $\rm arcsec$ to the outer boundary of $10^{2}$ $\rm arcsec$ ($N_{r_{\rm grid}}=10$). The angular grids $\theta_{\rm grid}$ and $\phi_{\rm grid}$ are sampled linearly between $0$ and $\pi /2$ ($ N_{\theta_{\rm grid}}=6, N_{\phi_{\rm grid}}=6$). This leads to 36 bins per radius and 360 bins in total.

\begin{table}
\centering
  \label{tab:Tmirror}
\begin{tabular}{ccc} 
	\hline \hline Position & $\hspace{2.5cm}$ All Orbit Families \\
	\hline
	$(x, y, z)$ & $\hspace{2.2cm}$ $(v_{x}, v_{y}, v_{z})$  &  \\
	$(x, y,-z)$ & $\hspace{2.3cm}$ $(v_{x}, v_{y}, -v_{z})$ &  \\
	$(-x,-y, z)$& $\hspace{2.3cm}$ $(-v_{x}, -v_{y}, v_{z})$ &  \\ 	
	$(-x,-y,-z)$& $\hspace{2.3cm}$ $(-v_{x}, -v_{y}, -v_{z})$ & \\
	\hline
\end{tabular}
\parbox{\columnwidth}{\caption{The Recipe of the Mirroring Scheme for
    All Types of Orbits in a Four-fold Symmetry Used for Modeling of
    Barred Galaxies. }}
\end{table}

\subsection{Weights of Orbits} \label{S:weighing_orbits}
The model constraints are the kinematic maps, usually including $V_o^l$ and $\sigma_o^l$ ($o$ stands for observation) in each aperture $l$, and Gaussian-Hermite (GH) coefficients $h_{3,o}^{l}$ and $h_{4,o}^{l}$, the surface brightness
in the 2D observational plane, and the 3D luminosity distribution deprojected from the 2D image. Note that $V_o^l$ and $\sigma_o^l$ are the parameters of GH function obtained by the full spectrum fitting; they are not the mean velocity and its dispersion unless all higher-order moments are zero.

The model is a superposition of thousands of orbit bundles, with each orbit bundle
$ k $ weighted by $ w_{k} $. We minimize the $\chi^2$ between data and
model to get the solution of the orbit weights. The $\chi^2$ is
contributed by two parts, the fitting to luminosity distribution and to kinematic maps:
\begin{equation}\label{NNLS}
\chi^{2}_{\mathrm{NNLS}}=\chi_{\operatorname{lum}}^{2}+\chi_{\mathrm{kin}}^{2}.
\end{equation}
\par
We allow relative errors of $1\%$ for 2D and 3D luminosity
distribution fittings. The 2D luminosity distribution $S_l$ is stored in the observational apertures in the observational plane. The contribution of orbit bundle $k$ in aperture $l$ is denoted as $S^{l}_{k} $. The 3D density distribution $\rho_n$ is stored in a 3D grid with 360 bins in total; the contribution of orbit bundle $k$ in bin $n$ is denoted as $\rho^{n}_{k}$. We thus have
\begin{equation}
\begin{split}
\chi_{\operatorname{lum}}^{2} &=\chi_{\mathrm{S}}^{2}+\chi_{\rho}^{2} \\
&= \sum_{l=1}^{ N_{\rm kin}}\left[\frac{\sum_{k} w_{k} S^{l}_{k}-S_{l}}{0.01 S_{l}}\right]^{2} +
\sum_{n=1}^{360}\left[\frac{\sum_{k} w_{k} \rho^{n}_{k}-\rho_{n}}{0.01 \rho_{n}}\right]^{2},
\end{split}
\end{equation}
where $N_{\rm kin}$ is the the number of apertures in one kinematic map ($N_{\rm kin}=476$ bins) and $w_k$ is the weight of orbit $k$.
\par
From observations, we describe the LOSVD profile $f_l$ in each aperture
$l$ as a GH distribution \cite[]{Gerhard.1993, Marel.1993} with parameters $(V_o^l, \sigma_o^l, h_{3,o}^l, h_{4,o}^l)$ and corresponding errors $(\Delta V_o^l, \Delta \sigma_o^l, \Delta h_{3,o}^l, \Delta h_{4,o}^l)$. 
When $V_o^l$ and $\sigma_o^l$ are chosen as the center and the width of the best-fitting Gaussian approximating the original LOSVD, this resulted in $h_{1,o}^l= h_{2,o}^l=0$.
The LOVSD contributions of orbit bundle $k$ at aperture $l$ we denote as $f_k^l$. If we expand $f^l_k$ in a GH series also with the central velocity and dispersion fixed at the observed $V_o^l$ and $\sigma_o^l$, then the resulting GH coefficients $h_{n, k}^{l}$ with $n=1,2,3,$ and $4$ will contribute linearly to the observations, so that
\begin{equation}
\chi_{\mathrm{kin}}^{2}=\sum_{l=1}^{ N_{\rm kin}} \sum_{n=1}^{n_{\rm GH}}\left[\frac{\sum_{k} w_{k} S_{k}^{l} h_{n, k}^{l} - S_{l} h_{n, o}^l}{S_{l} \Delta h_{n, o}^l}\right]^{2},
\end{equation}
where the model predictions are luminosity weighted in the same manner as the observations, and $n_{\rm GH}$ is the number of kinematic moments used for the fitting with here $n_{\rm GH} =4$. The errors of $(\Delta V_o^l, \Delta \sigma_o^l, \Delta h_{3,o}^l, \Delta h_{4,o}^l)$ are usually provided directly from observations, while we derive $\Delta h_{1,o}^l, \Delta h_{2,o}^l$ following \citet{Rix.1997}. The luminosity density is usually easy to fit, so that $\chi_{\mathrm{kin}}^{2}$ is the dominant term contributing to goodness of fit $\chi^{2}_{\mathrm{NNLS}}$ \cite[e.g.,][]{ling.1018}. 
We use the nonnegative least squares (NNLS) implementation \cite[]{Lawson.1974} to find the solution of orbit weights by minimizing the $\chi^2_{\mathrm{NNLS}}$ between data and model following \texttt{VdB08}.

\section{Application to a Mock Galaxy}\label{S:mock}

To test our model's ability of recovering the pattern speed, underlying mass profile, viewing angles, BP/X orbital structures, and internal orbit distribution, we apply our model to mock IFU data created from a simulated barred spiral galaxy. 

\subsection{Mock Data}
We use an \textit{N}-body barred galaxy model presented in \cite{Shen.2010}. This is a
Milky Way-like galaxy with a bar and spirals that recovers many
observed features of the Milky Way \cite[]{shen.2015}. The simulation
contains $10^{6}$ equal-mass particles, and the total stellar mass is
$M_{*} = 4.25 \times 10^{10} M_{\odot} $. 
A rigid logarithmic halo potential is adopted $\Phi=\frac{1}{2} V_{\mathrm{0}}^{2} \ln \left(1+r^2/R_{\mathrm{c}}^2\right)$, in which the scale velocity and scale radius are $V_{\mathrm{0}}=250  \hspace{.05cm} \mathrm{km}\mathrm{s}^{-1}$ and $R_{c}=15 \hspace{.05cm} \mathrm{kpc}$,  respectively. The bar forms at $t=2.3$ $\mathrm{Gyr}$, and we take a snapshot at $t=2.4$ $\mathrm{Gyr}$ right after the bar formation. The bar rotates with a pattern speed of $ \Omega_{ \mathrm{p}} \simeq 38 \hspace{.08cm} \mathrm{km \hspace{.04cm} s^{-1} \hspace{.04cm} kpc^{-1} }$ and has a half-length of $\simeq  \hspace{.03cm} 4 \hspace{.05cm} \mathrm{kpc}$ (corotation radius $  \simeq  \hspace{.03cm} 4.7 \hspace{.05cm} \mathrm{kpc} $). The end-to-end separation between the outer two edges of the $ \mathrm{X} $-shaped structures is	$\simeq \hspace{.03cm} 4 \hspace{.05cm} \mathrm{kpc} $ along the major axis and $\simeq  \hspace{.03cm} 2.4 \hspace{.05cm} \mathrm{kpc} $ along the vertical minor axis \cite[]{li.2012}. The effective radius of the galaxy is $R_e \simeq 3 \mathrm{kpc}$.

\par 
We adopt a distance of $41$\,Mpc so that 
$1\arcsec \simeq 200\,\mathrm{pc}$.
We project the simulation snapshot to the observational plane with an
inclination angle of $\theta_{\rm true} = 60 ^{\circ}$ and bar angle
of $\varphi_{\rm true} = -45^{\circ}$. Throughout the paper, we illustrate the creation of mock data, model fitting, and recovery with this version of projection ($\theta_{\rm true} = 60 ^{\circ}$ and $\varphi_{\rm true} = -45^{\circ}$). The results of another five sets of mock data with different projection angles are presented in the Appendix.
\par
\begin{figure} \label{error}
	\centering	%
	\includegraphics[width=\linewidth]{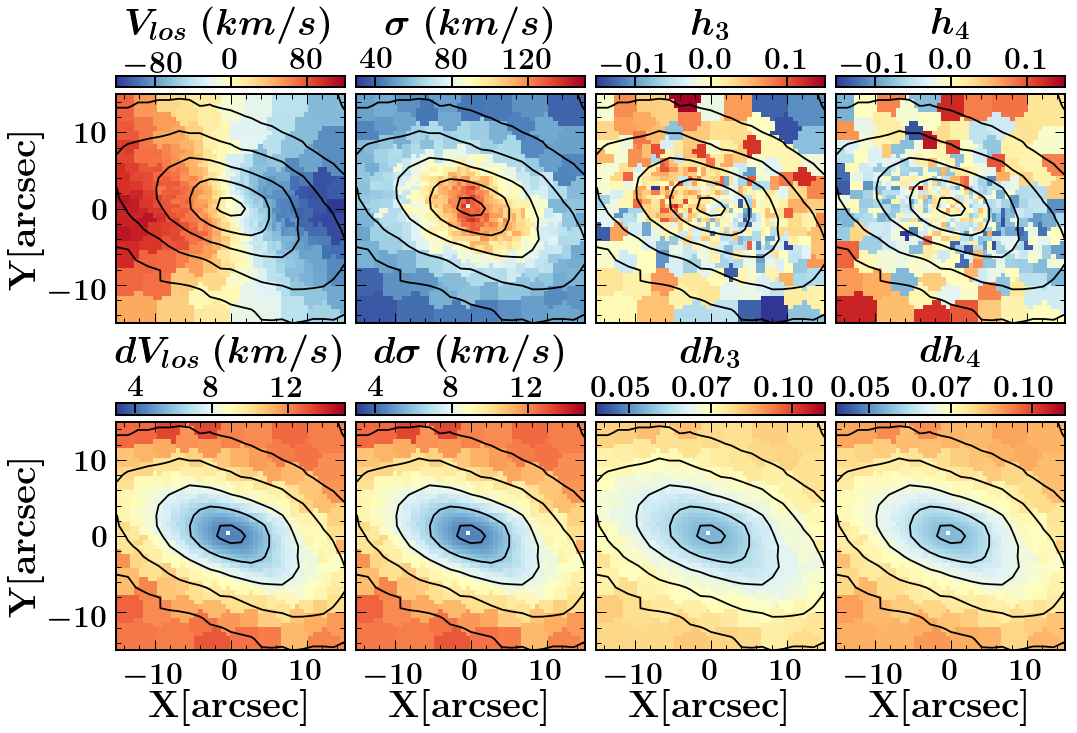}
	\hspace{8pt}%
	\caption{The mock kinematic maps (top) and error maps
          (bottom), overplotted with contours indicating the surface
          mass density. The panels from left to right are the mean
          velocity $V$, velocity dispersion $ \sigma $, the GH
          coefficients $h_{3}$, and $h_{4}$,  which are perturbed by their corresponding error maps. The mock data are created
          from a simulated bar galaxy, projected with the disk inclination
          angle $ \theta_{\mathrm{true}}=60^\circ$ and the bar angle $\varphi_{\mathrm{true}} = -45^\circ$. 
          }%
	\label{fig:nois}%
\end{figure}

The mock image is created with a spatial resolution of $ 1 $ arcsec pixel$^{-1}$ similar to what we have done in \cite{Behzad.2021}. To create mock kinematic data extended out to $1 \textit{R}_{e}$, we first separate particles into pixels of $1 \times 1\arcsec$, then apply Voronoi binning
\cite[]{Cappellari.2003} with the target signal-to-noise ratio
threshold of $S/N = 35$ (using number of particles over Poisson error) which leads to $N_{\rm kin}=476$ Voronoi bins for each kinematic map. Then we fit a GH profile to the LOSVD of particles in each bin, and obtain $(V_o^l,
\sigma_o^l, h_{3,o}^l, h_{4,o}^l)$ directly from the fitting.
To create mock kinematic errors, we use a logarithmic function inferred from the CALIFA data to construct the
errors \cite[]{Tsatsi.2015}. We then perturb the kinematic data by adding random noises inferred from the error maps to have realistic noisy data. The kinematic maps and the corresponding error maps for our mock galaxy are shown in Fig. \ref{fig:nois}.  

\subsection{Best-fitting Models}

\subsubsection{Exploring the Parameter Space}

\begin{figure*}
	\centering	%
	\includegraphics[width=2.0\columnwidth]{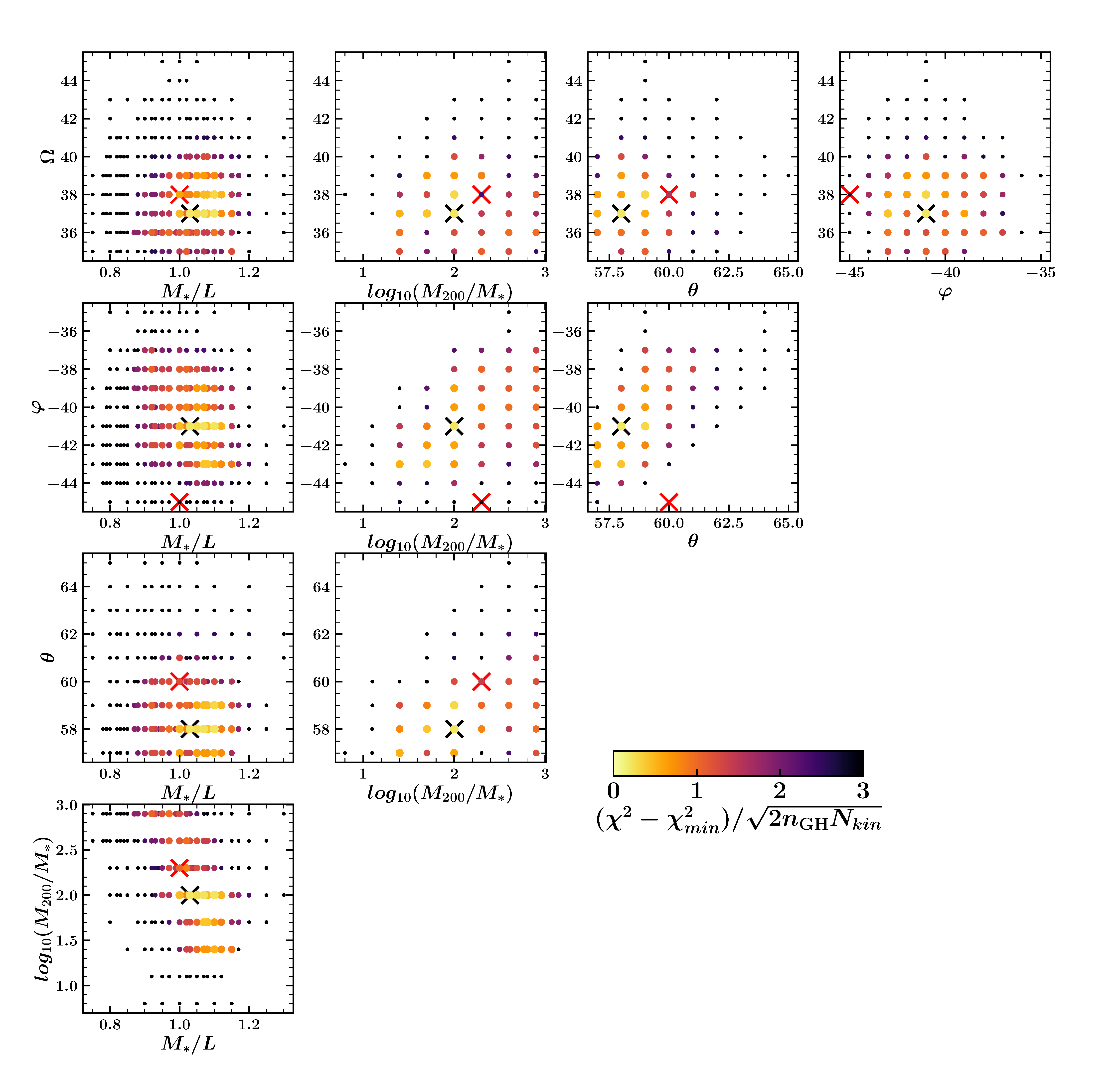}
	\hspace{8pt}%
	\caption{The parameter grid we explored for the model
          fitting. The five hyperparameters are stellar mass-to-light
          ratio $M_{*}/L$ in solar units, dark matter halo mass $\log
          M_{200}/M_{*}$ in unit of stellar mass, inclination angle of the disk $\theta$ in degrees, the bar angle with respect to major axis of the disk $\varphi$ in degrees, and the pattern speed in units of $\mathrm{km \hspace{.04cm}
            s^{-1} \hspace{.04cm} kpc^{-1} }$. Each point is one model color-coded
          according to their $\chi^{2}$ values shown in the color bar. 
          The points with $(\chi^{2}-\chi^{2}_{min})/\sqrt{2n_{\rm GH}N_{\rm kin}}<1$
          indicate models within $1 \sigma$ confidence level. The black crosses
          indicate the best-fitting model, while the red crosses denote the true values from the simulation (or those chosen for creating the mock data).
          }%
	\label{fig:chi2}%
\end{figure*}

\begin{figure*}
	\centering	%
	\includegraphics[width=1.7\columnwidth]{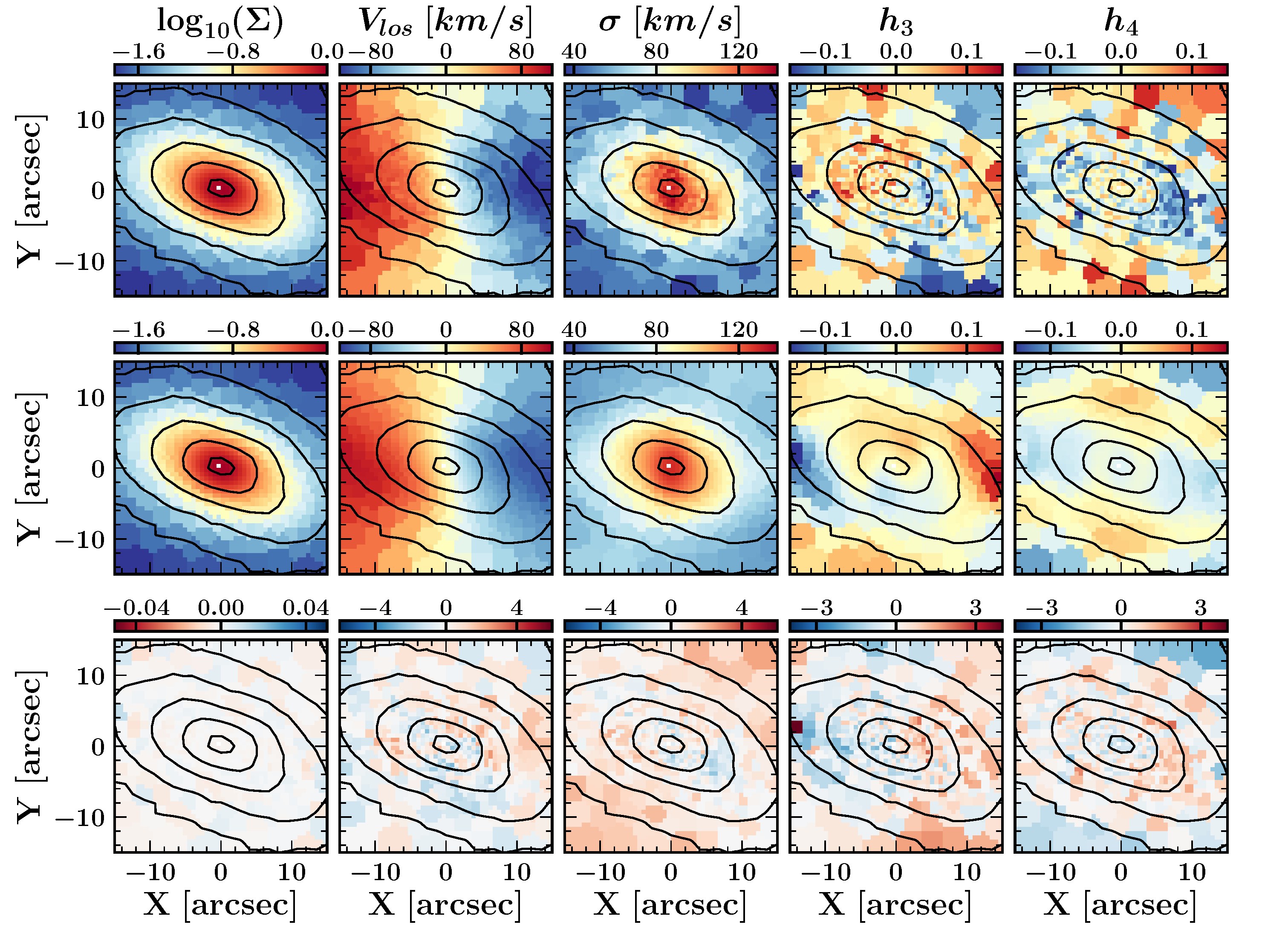}
	\hspace{8pt}%
	\caption{The best-fitting model of a mock barred galaxy. Columns from left to right represent
          the 2D surface density, velocity, velocity dispersion, $h_{3}$ and $h_{4}$. The first row shows the mock data within $1 \textit{R}_{e}$ created with $\theta_{ \rm true}=60^{\circ}$ and
          $\varphi_{ \rm true}=-45^{\circ}$. The second row shows the best-fit Schwarzschild model, obtained with $\theta=58^{\circ}$ and $\varphi=-41^{\circ}$. The third row shows the residuals, computed as the difference between the mock data and the model, divided by the uncertainties of mock data at each bin. Overplotted black contours indicate the surface mass density of the mock image.}%
	\label{fig:kin_map}%
\end{figure*}
Our model contains five free hyperparameters: constant stellar mass-to-light ratio $M_{*}/L$,  inclination angle $\theta$, bar azimuthal angle $\varphi$, bar pattern speed $\Omega$, and the DM virial mass $M_{200}/M_{*}$ as described in Section~\ref{S:pot}.
\par
For each  model, as described in Section~\ref{S:weighing_orbits}, we minimize the $\chi^2_{\rm NNLS}$ of GH coefficients $h_1,h_2,h_3,$ 
and $h_4$ for solving the orbit weights. After we have obtained the orbit weights of a model, we can then extract the model predicted mean velocity, velocity dispersion, $h_3$, and $h_4$ at each aperture, which leads to a direct comparison to the observed kinematic maps. It turns out that $\chi^2_{\rm lum}$ is very small and negligible ($\sim 1\%$) comparing to the fitting of kinematics ($\chi^2_{\rm kin}$), even for a barred galaxy. We evaluate goodness of fit between the model and observations as
\begin{equation}
  \label{eqn:chikin}
\begin{split}
\chi^{2}=\sum_{l=1}^{N_{k i n}}\left[\left(\frac{V_{m}^{l}-V_{o}^{l}}{\Delta V_{o}^{l}}\right)^{2}+\left(\frac{\sigma_{m}^{l}-\sigma_{o}^{l}}{\Delta \sigma_{o}^{l}}\right)^{2}+\right. \\
\left.\left(\frac{h_{3,m}^{l}-h_{3,o}^{l}}{\Delta h_{3,o}^{l}}\right)^{2}+\left(\frac{h_{4,m}^{l}-h_{4,o}^{l}}{\Delta h_{4,o}^{l}}\right)^{2}\right],
\end{split}
\end{equation}
where $ V_{m}^{l} $, $\sigma_{m}^{l}$, $h_{3,m}^{l}$, and
$h_{4,m}^{l}$ are the model predictions ($m$ stands for model), $ V_{o}^{l} $,
$\sigma_{o}^{l}$, $h_{3,o}^{l}$, and $h_{4,o}^{l}$ are observations with errors of $\Delta V_{o}^{l} $, $\Delta \sigma_{o}^{l}$, $\Delta h_{3,o}^{l}$, and $\Delta h_{4,o}^{l}$. In principle, the $\chi^{2}$ defined in this way should be strongly correlated with $\chi^{2}_{\mathrm{NNLS}}$. But there could be some difference caused by the the usage of $\Delta h_{1,o}^{l}, \Delta h_{2,o}^{l}$ analytically derived and sometimes significant numerical noise in $\chi^2_{\mathrm{NNLS}}$ introduced by the GH expansion of the LOSVD of each orbit bundle.
\par
We take an iterative process to search for the best-fitting model in the parameter grid, and $\chi^{2}$ defined in equation~(\ref{eqn:chikin}) is used to perform the iteration process.

We start with one initial trial of the hyperparameters. We then walk two steps in every direction of the parameter grid by taking relative large intervals of $ 0.05 $, $ 2 $, $ 2 $, $ 2 $, and $ 0.6 $ for $ M_{*}/L $, $ \Omega $, $ \theta $, $ \varphi $, and $\mathrm{log_{10}}(M_{200}/M_{*})$, respectively. Once the models are computed, we start an iterative process by selecting models with $\chi^{2} - \chi^{2}_{\rm min} < \sqrt{2} \Delta \chi^2$ from the existing models. We adopt $\Delta \chi^{2} \equiv \sqrt{2 n_{\rm GH} N_{\rm kin}}$ as our $ 1 \sigma $ confidence level, which is consistent with the $\chi^2$ fluctuation caused by numerical noise of the model  
\footnote{It was obtained by a bootstrapping process in the following: in a single model with fixed potential and orbit library, we perturb the kinematic data with its errors and fit the model to the perturbed data for many times. The standard deviation of $\chi^2$ obtained from these fittings are taken as the $\chi^2$ fluctuation caused by numerical noise of the model. In the classic statistic analysis for analytic models fitting to data, $1\sigma$ confidence level is determined by $\Delta \chi^{2}=1$. However, it is not suitable for our case where the model numerical noise is dominating the $\chi^2$.  The confidence level we adopt is not motivated by robust statistical consideration (more discussion on it could see \citet{Lipka2021}), but practically it works well in covering the true values in our model test.} \cite[]{ling.1018}.
The iteration will stop once the model with minimum $\chi^{2}$ is found and the models on the parameter grid around it are all calculated. Then, we halve the parameter step sizes to better sample the grids around the best-fitting models.

\par
We show the final parameter grid explored for modeling of the mock
galaxy in Fig \ref{fig:chi2}. Each point is one model colored by $(\chi^{2}-\chi^{2}_{\rm min})/\sqrt{2n_{\rm GH}N_{\rm kin}}$ value; note that
$\sqrt{2n_{\rm GH}N_{\rm kin}}$ represents the $1\sigma$ confidence level.  A
total number of $ \sim 7000 $ models are calculated, and parameter space around the best-fitting model is  well filled.
\par
We calculate the mean value of parameters from the models within the $
1 \sigma $ confidence level, and use the minimum and maximum values of
the parameters with these $1\sigma$ models as the lower and upper limits of the $ 1 \sigma $ error. We thus obtained $\theta = 58 \pm 3 ^\circ$, $\varphi = -41 \pm 3 ^\circ$, $\Omega =  37 \pm 3 \hspace{.08cm} \mathrm{km \hspace{.04cm} s^{-1} \hspace{.04cm}
kpc^{-1} }$, $M_{*}/L=1.03\pm 0.11$, and DM virial mass $\log_{10}(M_{200}/M_{*})=2.0 \pm 0.6$,  which generally recovers the true values of the simulation with $\theta^{\rm true} = 60^\circ$, $\varphi^{\rm true} = -45^\circ$, $\Omega^{\rm true} =  38
\hspace{.08cm} \mathrm{km \hspace{.04cm} s^{-1} \hspace{.04cm}
kpc^{-1} }$, $(M_{*}/L)^{\rm true}=1.0$, and $\log_{10}(M_{200}/M_{*})^{\rm true}= 2.3$. We have a large uncertainty on the DM virial mass due to limited data coverage. The bar azimuthal angle $\varphi$ is recovered less well, which might be caused by the degeneracy of the three viewing angles $(\theta_{\rm bar},\varphi_{\rm bar},\psi_{\rm bar})$. We have fixed the bar position angle $\psi_{\rm bar}$ by directly measuring it with respect the disk major axis ($\Delta \psi_j'$) in the image (see Section 2.1.1). This measurement could cause an uncertainty of a few degrees on $\Delta \psi_j'$, which could further cause the bias of a few degrees on $\varphi_{\rm bar}$.
\par

We show the best-fitting model in Fig~\ref{fig:kin_map}. The columns from left to right are
surface density, mean velocity, velocity
dispersion, $h_{3}$ and $h_{4}$, and the rows from top to bottom are mock
observational data, our best-fitting
model, and residuals. The residuals are calculated as the difference between the
data and the model, divided by the uncertainties of mock data at each
bin. The black contours indicate the surface density of the mock
image. Our model matches the data well, especially the main features of a barred galaxy, including the barred shape in surface
density map, the zero velocity curvature in the velocity map, the boxy shape in the $\sigma$ map, and two minima in $h_{4}$ located at the end of the barred structure. 

\subsubsection{The Bar Pattern Speed}

\begin{figure}
	\centering	%
	\includegraphics[width=\columnwidth]{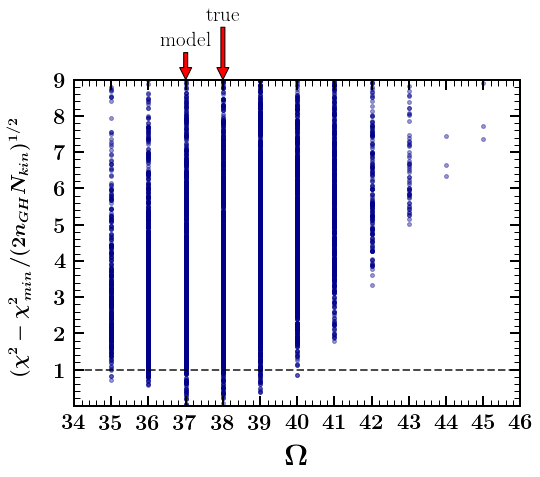}
	\hspace{8pt}%
	\caption{Normalized $ \chi^{2} $ as functions of $\Omega$. Each point indicates a model. The vertical red arrows mark the model and true values of $\Omega$. }%
	\label{fig:chi2_om}%
\end{figure}

The marginalized $\chi^2$ as a function of the pattern speed $\Omega$
is shown in Fig. \ref{fig:chi2_om}. The models with
$(\chi^{2}-\chi^{2}_{\rm min}) < \sqrt{2n_{\rm GH}N_{\rm kin}}$ are chosen as models
within the $ 1 \sigma $ confidence level. We obtained the pattern
speed of $\Omega =  37 \pm 3 \hspace{.08cm} \mathrm{km \hspace{.04cm}
  s^{-1} \hspace{.04cm} kpc^{-1} }$ from our models. While the true
pattern speed of the simulation is $\Omega_{\rm true} =  38
\hspace{.08cm} \mathrm{km \hspace{.04cm} s^{-1} \hspace{.04cm}
  kpc^{-1} }$, which is thus well recovered with a $1\sigma$ significant relative
uncertainty of $\sim 10\%$. The recovery of bar pattern speed works similarly well for galaxies with a wide range of inclination angles and bar angles (see Figure~\ref{fig:kin_mapall} in the Appendix). 

A simple and model-independent method for measuring pattern speed $
\Omega $ of external barred galaxies is introduced by
\cite{Tremaine.1984} (TW method). It uses the one-dimensional profiles of surface brightness $\Sigma (x)$ and LOS velocity $V_{los}$ measured along the bar, with the coordinate of $x$  integrated from $- \infty$ to $\infty$ along a slit.
The TW method is widely used; however the accuracy of this approach depends on accurate determination of the disk position angle. It can lead to errors of 10\% (up to 100\%) for $\Omega$ with inaccuracies of a few degrees in the disk position angle \cite[]{Debattista.2003, Zou.2019}. Our approach hopefully will provide an independent way of determining the pattern speed of barred galaxies, consistent with the results from \cite{Vasiliev.2019c}.

\subsubsection{Enclosed Mass Profile}
\begin{figure}
	\centering	%
	\includegraphics[width=\columnwidth]{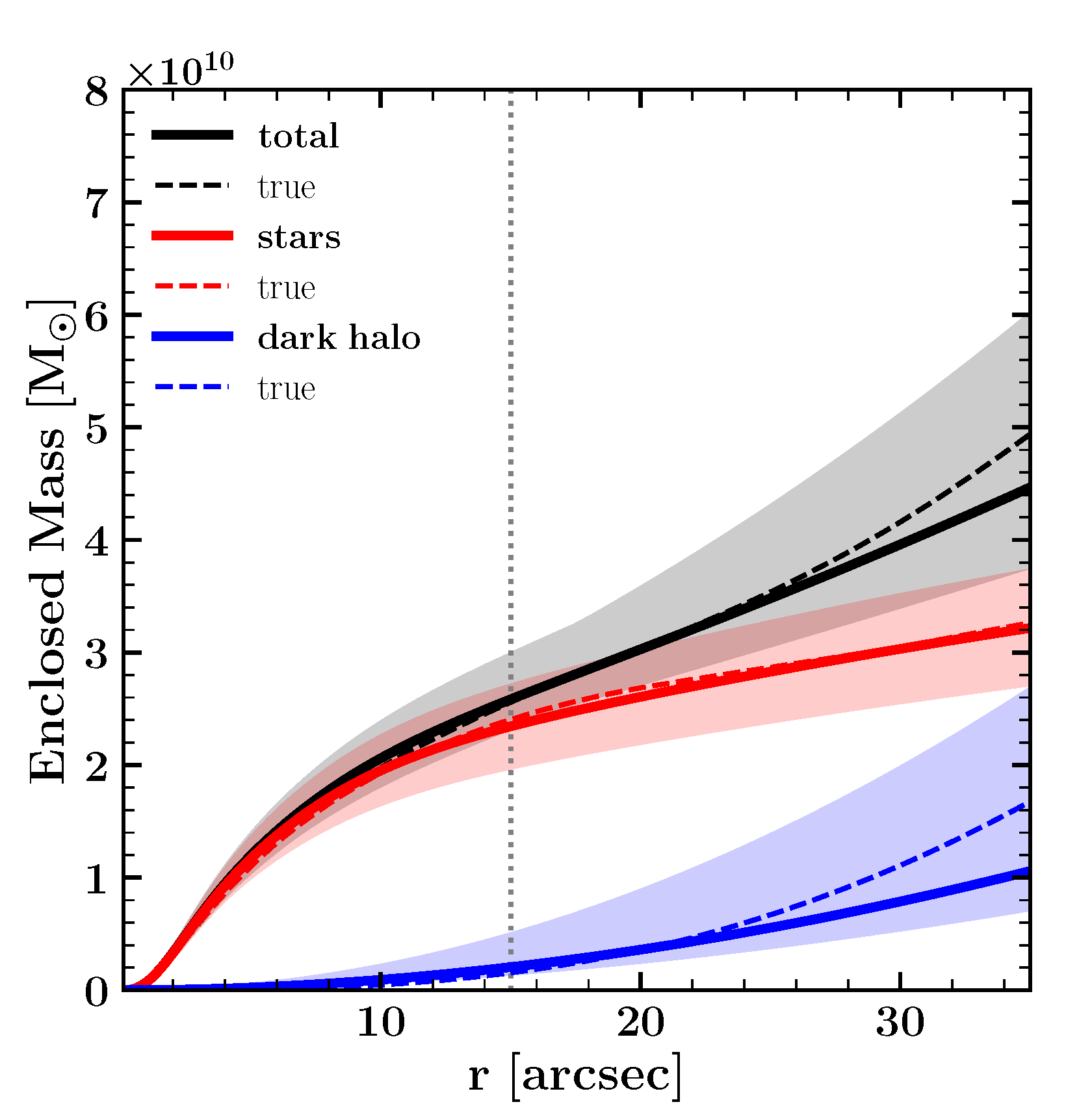}
	\hspace{8pt}%
	\caption{Enclosed mass profiles of the best-fitting model (solid lines) compared to those from the simulation (dashed lines). The red, blue, and black curves represent the stellar mass, dark matter mass, and total mass, respectively. The shaded regions indicate the $1\sigma$ uncertainty from our model. The vertical gray dotted is $\textit{R}_{e}$ and is indicative of the kinematics data extent. 
	}%
	\label{fig:mass}%
\end{figure}

Our model recovers well the underlying mass profiles as shown in Fig.~\ref{fig:mass}. The solid red, blue, and black lines indicate the stellar, dark matter, and total enclosed mass profiles. The filled regions indicate the $1 \sigma$ scatter of mass profiles of all models in the $1 \sigma$ confidence level. The true enclosed mass profiles of the simulated galaxy are plotted with the dashed lines of the same colors. 
We take all particles in the simulation with equal mass and luminosity when creating the mock data; thus the true stellar mass-to-light ratio should be unity. We found $M_*/L=1.03\pm0.11$ from our model, consistent with the true value. The stellar mass profile is well recovered.  
The simulation actually has a logarithmic DM halo, while we blindly fit it with an NFW halo in our model. We still find models generally matching the true DM mass profile, because we only focus on the inner $1R_e$ where the DM is a small fraction and our data constraint on the DM model itself is limited. Mass profiles are recovered similarly well for galaxies with different projection angles (see Figure~\ref{fig:kin_mapall} in the Appendix).

\begin{figure*}
	\centering	%
	\includegraphics[width=2\columnwidth]{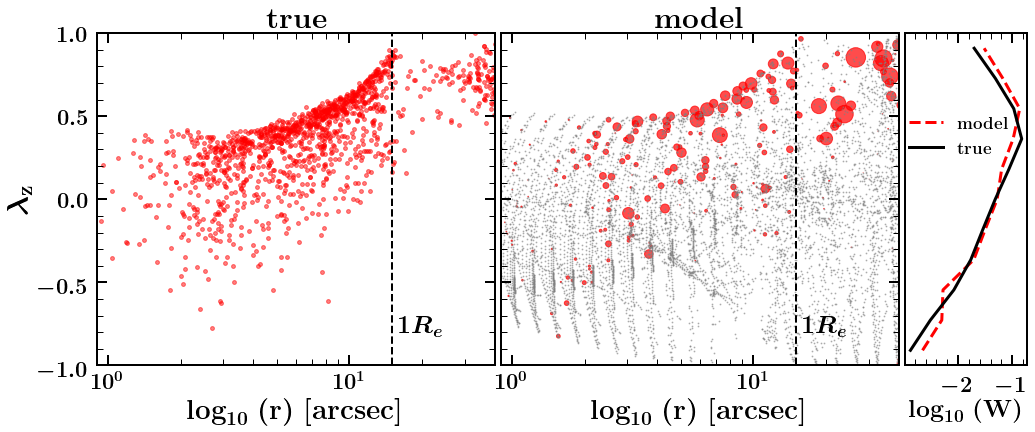}
	\hspace{8pt}%
	\caption{Stellar orbit distribution in the phase space of circularity $\lambda_{z}$ vs. time-average radius $r$. The left panel shows the true distribution in the simulation with 2500 randomly selected orbits. The middle panel show the distribution from our best-fitting model, with the larger size of the red dots representing orbits with higher weights from the minimum value of $10^{-8}$ to a maximum of $10^{-1}$ (the total weight of all orbits is $=1$). The gray dots are orbits sampled but with zero weight in the model. The vertical dashed line represents $1\textit{R}_{e}$ which indicates the maximum radius covered by our kinematic data. In the right side panel, we compare the $\lambda_z$ distribution as a function of $\rm \log_{10}$ of the weight for all the orbits at $r \leq R_e$, in the simulation (black solid curve) and in our best-fitting model (red dashed curve).
	}%
	\label{fig:Cir}%
\end{figure*}

\subsubsection{Orbital Circularity Distribution} \label{sec:cir}
We describe the stellar orbit distribution as the probability density of orbits in the  space of radius $r$ versus circularity $\lambda_z$, defined as the angular momentum $L_z$ normalized by the maximum that is allowed by a circular orbit with the same binding energy. In the orbit-superposition model, $\lambda_z$ and $r$, are taken as the average of an orbit stored with equal time steps. We calculate $\lambda_z$ as \cite[]{ling.1018}:
\begin{equation}
\lambda_{z}=\overline{L_{z}} /\left(\bar{r} \times \overline{V}_{\mathrm{rms}}\right),
\end{equation}
where $\overline{L_{z}}=\overline{x v_{y}-y v_{x}}, \quad \bar{r}=\overline{\sqrt{x^{2}+y^{2}+z^{2}}}$ and $\overline{V}_{\mathrm{rms}}^2 = \overline{v_{x}^{2}+v_{y}^{2}+v_{z}^{2}+2 v_{x} v_{y}+2 v_{x} v_{z}+2 v_{y} v_{z}}$. Nearly circular orbits have $\lambda_{z} \sim 1$, while for box orbits the time-averaged angular momentum vanishes so that $\lambda_{z} \sim 0 $. 

\par
For the particles in the simulation, we randomly select $2500$ particles and integrate their orbits in the frozen \textit{N}-body potential using AGAMA \footnote{\url{https://github.com/GalacticDynamics-Oxford/Agama}} \cite[]{Vasiliev.2019}.  We freeze the \textit{N}-body system at the given snapshots, then calculate potentials from the particle distribution using multipole expansion of spherical harmonics \cite[]{Binny.2008}, and we add a rigid logarithmic DM halo. Orbits are computed in a corotating frame with the true pattern speed of $ \Omega_{ \mathrm{p}} \simeq 38 \hspace{.08cm} \mathrm{km \hspace{.04cm} s^{-1} \hspace{.04cm} kpc^{-1} }$.  Once the orbits are computed, then $r$ and $\lambda_z$ are calculated from average of particles sampled from an integrated orbit similar to that in the orbit-superposition model. 

In Fig. \ref{fig:Cir}, we show the stellar orbit distribution in $r$ versus $\lambda_z$ for the simulation in the left panel, and for our best-fitting model in the right panel. The size of circles indicate the orbits weights from $w_{i}=10^{-8} $ to $w_{i}=10^{-1}$. Our model generally recovers the true circularity distribution. The data constraints only cover the inner $1\,R_e$, where the bar-trapped orbits are dominating. At $r>1\,R_e$, the model is only constrained by the density distribution that is dominated by the disk. Even so, the corresponding disk orbits with high circularity in the model are still similar to that in the simulation. 

\subsection{Orbital Decomposition}\label{S:ron_alz}
The stellar orbit distribution in $\lambda_{z}$ versus $r$ cannot fully reveal the complicated orbital structures of the bar. 
In this section, we use the orbital frequency analysis to generate a detailed view of the orbital structures in our model and compare it to the true in the simulation.
 
\subsubsection{Orbital Frequency Analysis}

\begin{figure*}
	\centering	%
	\includegraphics[width=2\columnwidth]{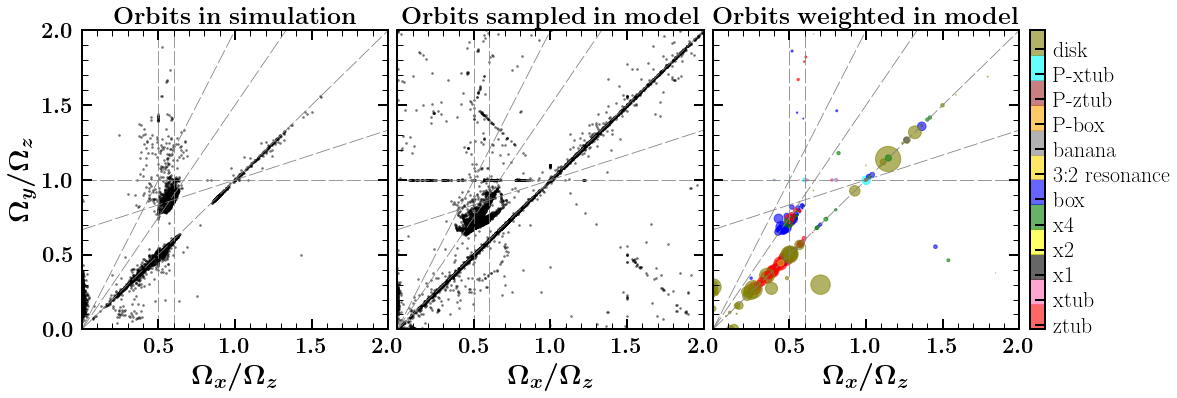}
	\hspace{8pt}%
	\caption{ Stellar orbit distribution in the frequency map of $\Omega_x/\Omega_z$ vs. $\Omega_y/\Omega_z$. From left to right, the three panels are orbits in simulation, all orbits sampled in our model, and the orbits weighted in our best-fitting model. In the right panel, the larger size of the circles indicates a higher orbit weight from a minimum value of $10^{-8}$ to a maximum of $10^{-1}$, colors indicate the different orbit types from NAFF autoclassification. Dashed lines mark some common resonant sequences. 
	}%
	\label{fig:freq}%
\end{figure*}

Frequency analysis is a key tool to understand the orbital structures
of barred galaxies, especially in  
recognizing the resonant orbits.
We use the NAFF software \footnote{\url{https://bitbucket.org/cjantonelli/naffrepo/src/master/}}  \cite[]{Valluri.1998, Valluri.2016} to compute the fundamental frequencies for the orbits, and compare the true distribution in the simulation with that in our model.
\par 
In our orbit-superposition model, a total number $n_{E} \times n_{R} \times n_{\phi} \times 3^{n_{\text{\rm dith}}} \times n_{\text{orbit library}} $ =
$40 \times 20 \times 10 \times 3^{3} \times 2 = 432,000 $ are sampled, and subsequently every $3^3$ dithered orbits are considered as an orbital bundle.
We subsequently calculate the frequencies for all $16,000$ central orbits of each bundle. For particles in the \textit{N}-body simulation, we integrate $16,000$ randomly selected orbits in a frozen potential as described in Section~\ref{sec:cir}. We then compute the fundamental frequencies in Cartesian coordinates for each orbit. 
\par
Figure~\ref{fig:freq} shows the resulting orbit distributions in $\Omega_x/\Omega_z$ versus $\Omega_y/\Omega_z$, for the simulation in the left panel and for the model in the middle panel. The right panel only shows the orbits with nonzero weights in the best-fitting model, with the larger circle size indicating higher orbit weights from $w_{i}=10^{-8} $ to $w_{i}=10^{-1}$, normalized such that their sum is unity. 
The orbits we sampled in the model generally span the whole area in the $\Omega_x/\Omega_z$ versus $\Omega_y/\Omega_z$ map covered by the orbits from the simulation. This shows that, even though the orbits are launched only from the $x-z$ plane, this initial sampling is adequate. There is a small frequency shift for the clump of box
orbits, which could be caused by an imperfect match of vertical structures in the simulation by our deprojected 3D density distribution model.

\subsubsection{Bar Orbits Classification}
The NAFF software \citep{Valluri.2016} classifies the orbits into different types based on the frequency analysis. In this approach, disk orbits are identified as those orbits with apocenter radii larger than the  half-length of the bar ($r_{\mathrm{apo}} > 4 \hspace{0.05cm} \mathrm{kpc} $ for our simulated galaxy). Then the orbits within the bar are classified into z-tube, periodic z-tube, x-tube, periodic x-tube, box, periodic box, $x_{1}$, $x_{2}$, $x_{4}$, banana ($1:2$ resonance), and pretzel/fish orbits ($3:2$ resonance) orbits. We refer to \cite{Valluri.2016} for the details of automated classification of orbits.
\par
The colors in the right panel of  Fig. \ref{fig:freq} indicate the orbits types that are classified by NAFF. In our final model, the different types of orbits are weighted in a similar manner as that in the simulation.

\subsubsection{Recovery of BP/X Structure}
\begin{figure*}
	\centering	%
	\includegraphics[width=2\columnwidth]{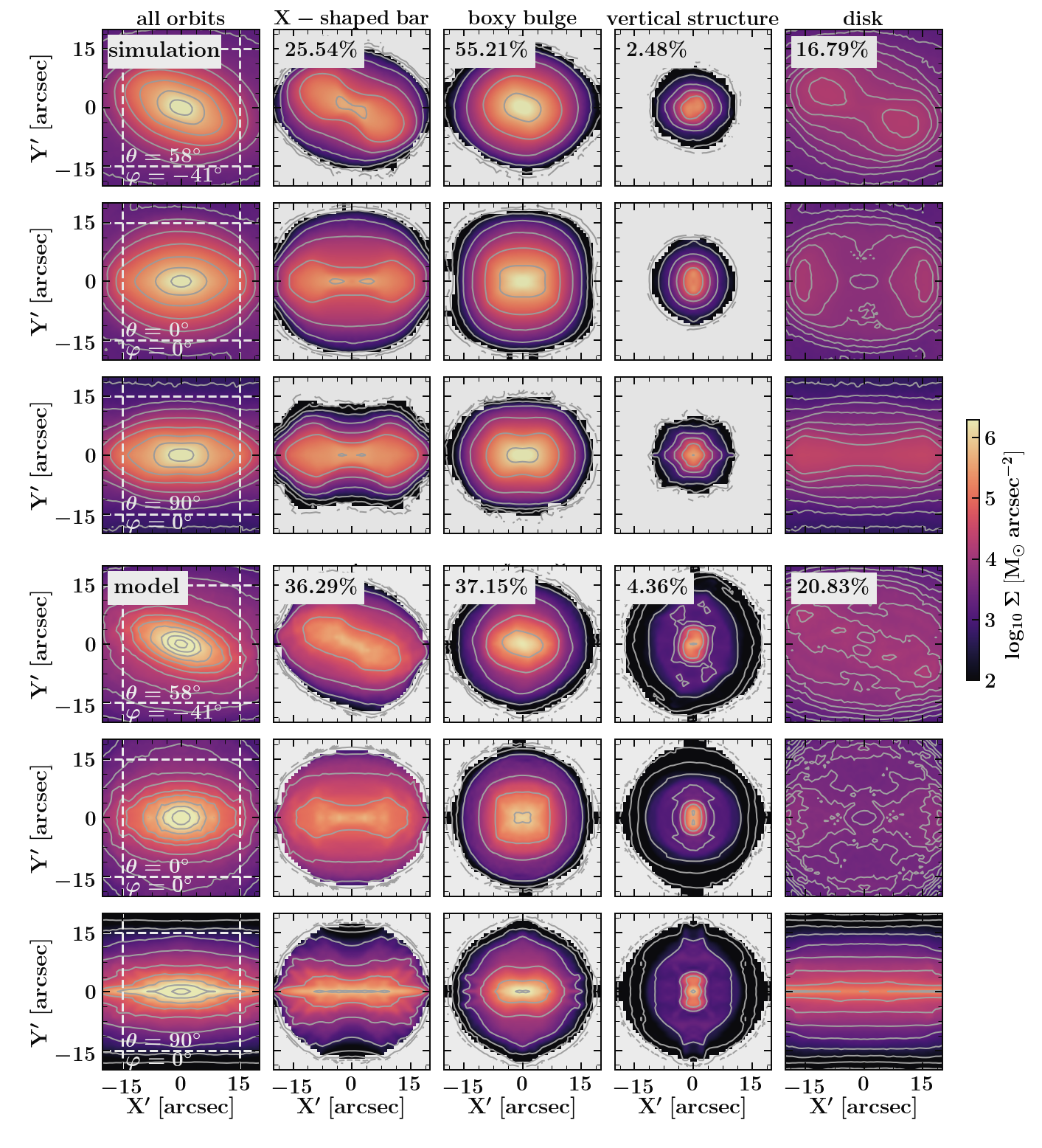}
	\hspace{8pt}%
	\caption{Projected surface density of different orbital structures from the simulation (top 3 rows) and from our best-fitting model (bottom 3 rows). Each 3  rows from top to bottom are projections with the galaxy orientated as observed (top), face-on (middle) and edge-on (bottom) views. Columns from left to right show the reconstructed surface densities of all orbits, the X-shaped bar, the boxy bulge, the vertically extended structure, and the disk orbits, respectively. The white dashed lines in the first column indicate the data coverage. The luminosity fraction of each structure within the the data coverage are indicated in the top row (in \%). }%
	\label{fig:orb_dec}%
\end{figure*}

In this section, we perform orbital decomposition to compare the morphology of different structures in our model to those in the simulation. The decomposition is based on the orbit types from the NAFF autoclassification.
\par
Guided by previous studies \cite[]{Portail.2015, Abbott.2017, Parul.2020} and our analysis of orbital classes in BP/X structures of a few simulations (Tahmasebzadeh et al. 2022, in preparation), we broadly divide all orbits into four groups: \textbf{(1) X-shaped bar}: including $x_{1}$, banana ($1:2$ resonance), z-tube, and periodic z-tube orbits as defined by NAFF, which are prograde short-axis tube orbits elongated along the bar and generate a pronounced X-shaped structure in edge-on and face-on projected images; \textbf{(2) boxy bulge}: including pretzel/fish orbits ($3:2$ resonance), periodic and nonperiodic box orbits that generate the boxy-shape structure in the face-on and edge-on images, and a faint X-shaped structure in which X-wings cross the center of the disk plane;
\textbf{(3) vertically extended structure}: a structure perpendicular to be bar constructed by x-tube, periodic x-tube, and $x_{4}$ orbits;
\textbf{(4) disk}:  all orbits that have apocenter radii larger than the half-length of the bar. 
Note that theoretically there might be another type of orbit, $x_{2}$, perpendicular to the bar. But we do not find any $x_{2}$ orbits, neither in our model nor in the simulation.
\par
We then reconstruct the 3D density distribution of each structure by summing the particles sampled from the orbits in each group.
For orbits in our best-fitting model, the orbit classification is based the central orbit of each orbital bundle with $3^3$ dither orbits bounded together,  we use particles sampled from all orbits and take the orbits in the same bundle following the orbit classification as the central orbits.
\par
In Fig. \ref{fig:orb_dec}, we show the surface densities of different structures by projecting them with the orientation as the galaxy was observed, with face-on and with edge-on views, respectively.  Columns from left to right are surface densities created by all orbits, X-shaped bar, boxy bulge, vertically extended structure, and disk orbits, respectively. With this orbital decomposition, the barred structures are well separated from the disk. 

In the projected view, the morphology of the whole galaxy, and the different structures, including X-shaped bar and boxy bulge, in our model match well those in the simulation.
Note that we do not have an X-shaped structure in the input 3D density distribution, where the barred bulge is mostly prolate, deprojected from the 2D image. Although an X-shaped structure is thus not explicitly included in the gravitational potential, the model can still support orbits leading to the X-shaped structure \citep[see also][]{Behzad.2021}. 

In the edge-on view, we can see that there are some mismatches between the intrinsic 3D shape of our model and the simulation. There are some very thin box orbits in our model, whereas these kinds of thin orbits are actually rare in the simulations. The mismatches of intrinsic 3D shape are caused by the lack of information of the true 3D density distribution with only a 2D image. Our 3D density distribution deprojected from the 2D image cannot reveal the edge-on structure perfectly when the galaxy is observed at a moderate inclination angle of $60^o$. 
Lacking of a perfect 3D density distribution, in the current model, we used a coarse 3D grid to record the 3D density distribution in the model fitting (see section 2.2), which does not aim for capturing all the fine structures (see figure~\ref{fig:3d_dens} in the Appendix). Thus, the residual from fitting of density distribution is still very small in our model.
\par 
We quantitatively compare the structures in our model to the simulation by calculating the luminosity fraction of different components within the data coverage. The luminosity fractions of different components in the simulation are $25.5\%$ X-shaped bar, $55.2\%$ boxy bulge, $2.5\%$ vertically extended structure, and $16.8\%$ disk, while for our model they are $36.3\%$ X-shaped bar, $37.1\%$ boxy bulge, $4.4\%$ vertically extended structure, and $20.8\%$ disk. Even though our model recovers the contribution of the different components rather well, this might be improved by introducing physically motivated constraints on the intrinsic shape of the bar to better match the 3D intrinsic shape. In this case, we need to modify the 3D grid recording the 3D density distribution to capture the fine structures.

\section{Conclusions}\label{S:con}
We modify the triaxial Schwarzschild model from \texttt{VdB08} to explicitly include a rotating bar. The gravitational potential is generated from a combination of a 3D stellar luminosity density multiplied with a stellar mass-to-light ratio and a spherical dark matter halo.
We use a two-component deprojection method to obtain the 3D stellar luminosity density from an observed 2D image through a combination of an axisymmetric disk and a triaxial (mostly prolate) barred bulge. We consider figure rotation of the galaxy
and corresponding gravitational potential with the bar pattern speed as a free parameter. We perform orbit integration in the bar corotating frame. But the initial starting points are sampled in inertial frame, and the kinematics $v_x,v_y,$ and $v_z$ in the inertial frame are stored, which is convenient for the fitting to data. 
We solve for the orbit weights of the model by simultaneously fitting the 3D luminosity density, the 2D surface brightness, and the stellar kinematic maps. We validate the method by testing it on mock IFU stellar kinematic maps generated from a simulated barred galaxy with a BP/X-shaped bulge.
Our model matches the observational data reasonably well, including the major properties of the bar in the observed 2D image and the stellar kinematic maps. 
The main results are as follows:
\par
(1) The disk inclination angle $\theta_{\rm disk}$ and the azimuthal angle of the bar in the disk plane $\varphi_{\rm bar}$ are well recovered with $1\sigma$ uncertainty of $\sim 3^\circ$. The bar pattern speed $\Omega$ is well recovered with a $1\sigma$ relative uncertainty of $\sim 10\%$ for galaxies in a wide range of projection angles.

\par
(2) The enclosed stellar and dark halo mass profiles are recovered with a relative uncertainties of $\sim 10\%$ within the data coverage.
 
\par
(3) We decomposed the galaxy into different orbital structures based on orbital frequency analysis. Based on the NAFF autoclassification, we confirm that all orbits supporting the barred structures in the simulation are indeed included in our model. The BP/X structures, including an X-shaped bar and a boxy bulge, in the simulation are generally recovered by our model

\par
We demonstrated that our model works well in reproducing the stellar kinematic properties of a barred galaxy and in uncovering subsequent key properties of barred galaxies like the bar pattern speed and the internal BP/X-shaped orbital structure. 
Due to the limited information on the intrinsic 3D shape of the barred bulge from the observed 2D image, the inference on the intrinsic shape of the bar is imperfect. This might be improved by introducing physically motivated priors on its intrinsic shape.

\section*{Acknowledgements}
We thank Eugene Vasiliev and Monica Valluri for useful discussions. The research presented here is partially supported by the National Key R\&D Program of China under grant No. 2018YFA0404501; by the National Natural Science Foundation of China under grant Nos. 945271001, 12025302, 11773052, 11761131016; by the ``111'' Project of the Ministry of Education of China under grant No. B20019; and by the Chinese Space Station Telescope project, and by the Deutsche Forschungsgemeinschaft under grant GZ GE 567/5-1(OG); and CAS Project for Young Scientists in Basic Research under grant No. YSBR-062. This work made use of the Gravity Supercomputer at the Department of Astronomy, Shanghai Jiao Tong University, and the facilities of the Center for High Performance Computing at Shanghai Astronomical Observatory. B.T. acknowledges support from  CAS-TWAS President's Fellowship for international PhD students, awarded jointly by the Chinese Academy of science and The World Academy of Sciences. G.v.d.V. acknowledges funding from the European Research Council (ERC) under the European Union's Horizon 2020 research and innovation programme under grant agreement No. 724857 (Consolidator Grant ArcheoDyn).
\\
\\
\textit{Software}: \textit{Agama} \cite[]{Vasiliev.2019}, \textit{NAFF} \cite[]{Valluri.1998, Valluri.2016}, \textit{Jupyter Notebook} \cite[]{Kluyver.2016},  \textit{matplotlib} \cite[]{Hunter.2007}, \textit{numpy}\cite[]{Harris.2020},  \textit{scipy}\cite[]{Virtanen.2020}.  

\bibliography{1.bibtex}{}
\bibliographystyle{aasjournal}

\appendix
\section{Model tests for mock galaxies with different projections} \label{app:other_mocks}
\begin{figure*}
	\centering	%
	\includegraphics[width=1.0\columnwidth]{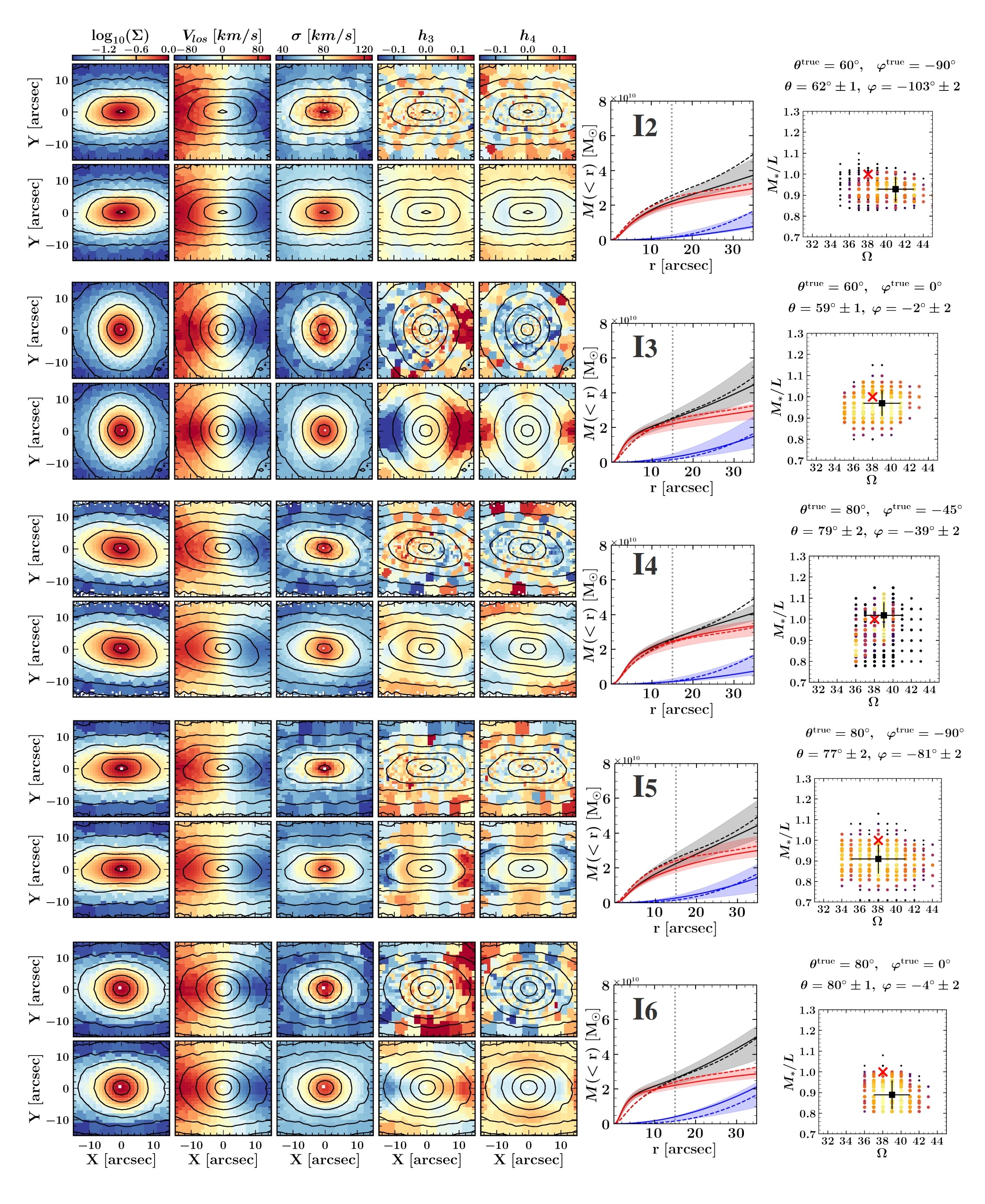}
	\hspace{8pt}%
	\caption{The best-fitting model of mock barred galaxies with different viewing angles of
           $\theta^{true}$ and $\varphi^{true}$, $I{2}$, $I{3}$, $I{4}$, $I{5}$, $I{6}$ from top to bottom. For each galaxy, we show the best-fitting kinematic maps (similar to Figure~\ref{fig:kin_map}), enclosed mass profile (similar to Figure~\ref{fig:mass}), and recovery of pattern speed $\Omega$ (similar to the top-left panel of Figure~\ref{fig:chi2}). In each panel of the right column, the red cross indicate the true value, the black dot with error bar indicate our best-fitting model with $1\sigma$ uncertainty.}%
	\label{fig:kin_mapall}%
\end{figure*}

We applies the method to a few more mock galaxies $I{2}$, $I{3}$, $I{4}$, $I{5}$, $I{6}$  created from the same simulation but with different inclination angle $\theta^{\mathrm{true}}$ and bar angle $\varphi^{\mathrm{true}}$, as listed in Table 1 of \cite{Behzad.2021}.
The major results are shown in Fig. \ref{fig:kin_mapall}. For each mock galaxy, we show the best-fitting model, the recovery of enclosed mass profile, and the pattern speed $\Omega$.
Generally, we fit the kinematic maps well, and recovery the model parameters well, including pattern speed and mass profile similarly well. 

\begin{figure*}
	\centering	%
	\includegraphics[width=1.0\columnwidth]{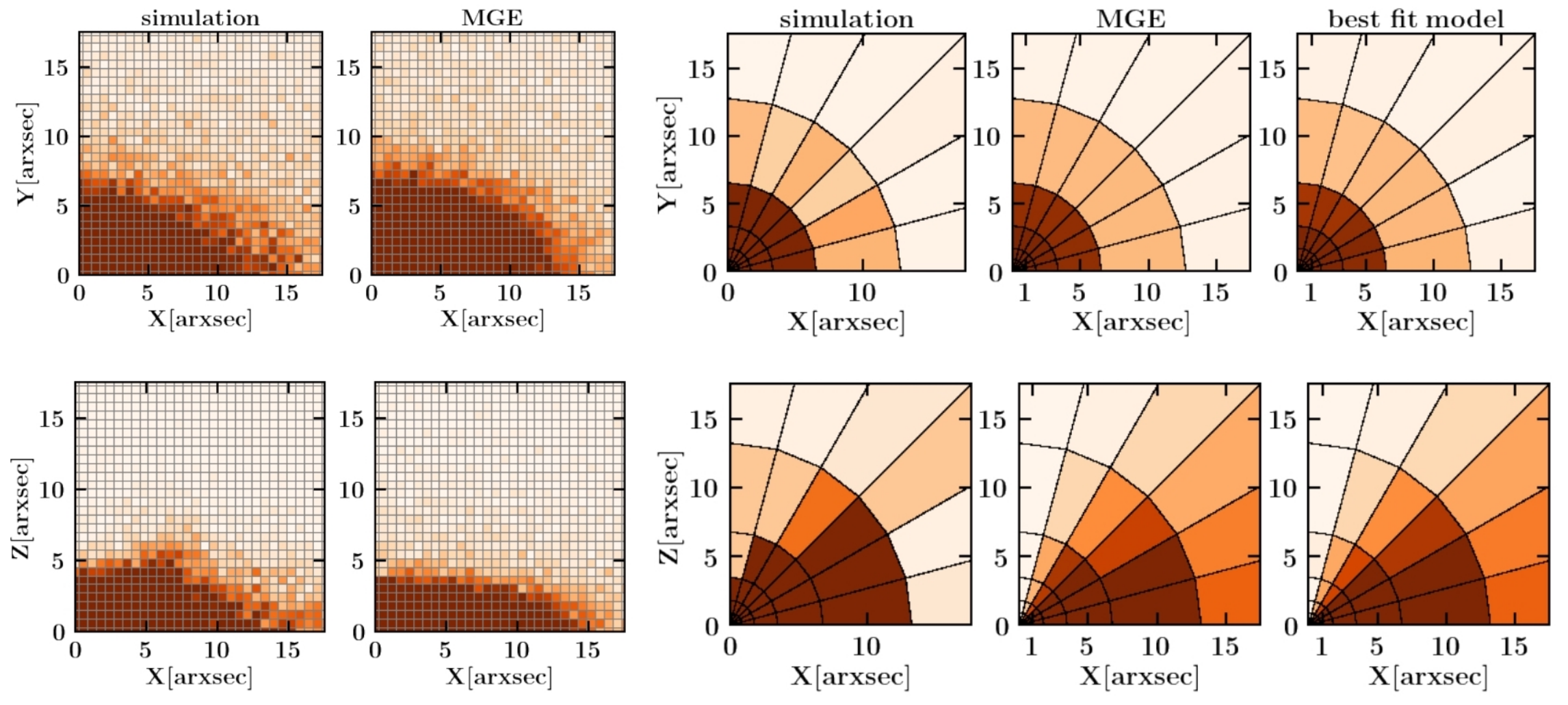}
	\hspace{8pt}%
	\caption{Displaying of 3D density distribution. The top and bottom rows show density distribution near $x-y$ and $x-z$ plane. The first two columns are the density distribution from the simulation and our MGE model, shown in fine 3D grid, the rest three columns are the density distributions from simulation, our MGE model, and our best-fitting model, shown in the coarse 3D grid adopt in our model. The difference of density distributions in different models are washed by the coarse 3D grid recording them. The residual between our best-fitting model and the MGE model put in as 3D density constraint is very small. }%
	\label{fig:3d_dens}%
\end{figure*}

\clearpage

\end{document}